\documentstyle[epsfig]{aipproc2}
\begin{document}
\title{Introduction to Diffractive Photoprocesses}

\author{Graham Shaw$^*$}
\address{$^*$Department of Physics and Astronomy, University of Manchester, 
Manchester M13 9PL, U.K.}

\maketitle

\begin{abstract}
The objectives of my talk are to provide a very brief introduction to 
diffractive photoprocesses 
in general and the colour dipole model in particular; and to comment on
possible gluon saturation effects at HERA and beyond.

\end{abstract}

\section*{Introduction}
\begin{figure}[htb]
\begin{center}
\rotatebox{270}{\psfig{file=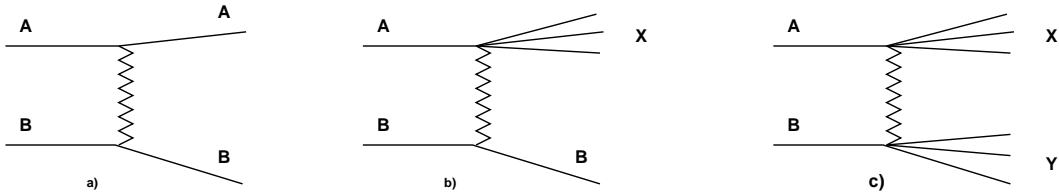,height=14cm}}
\vspace{0.3cm}
\caption{(a) elastic (b) singly dissociative and (c) double dissociative
diffractive processes. (Figure from [1]).
\label{fig:had_dif}
}
\end{center}
\end{figure}

Diffraction exchange is the study of vacuum exchange at high energies. It is
frequently divided into elastic, singly-dissociative and double dissociative 
processes as illustrated in Figure \ref{fig:had_dif}, where $A$ and $B$ may be photons or 
hadrons and $X$ and $Y$ may be single particles or an inclusive sum 
over $ n \ge 1$  particle states. The
wiggly line indicates
an exchange of energy and momentum, but no non-zero colour or flavour quantum
numbers may be exchanged. High energy means that the square of the centre of
mass energy $s = W^2$ is much larger than any other energy scale:
$$
s \gg t, m_X^2, \ldots \; \; .
$$
For diffractive processes initiated by virtual photons, the latter include the
virtuality $Q^2$, implying
$$
x = Q^2/s \ll 1   \; \; .
$$
We note that $t, m_X^2, Q^2$ can themselves become large, provided they remain 
much smaller than $s$.

Experimentally, diffractive processes are characterized by two distinctive
features: {\bf rising
cross-sections} and {\bf rapidity gaps}. The two groups of final state 
particles in Figure \ref{fig:had_dif} emerge in roughly the forward and backward directions
in the centre of mass frame; and are well-separated in rapidity or
pseudo-rapidity
$$
\eta = - \ln \tan(\frac{\theta}{2})
$$
where $\theta$ is the polar angle with respect to the beam direction.
Such rapidity gaps are characteristic of colour singlet exchange, in 
contrast to the hadronization strings associated with colour exchange. 
They occur not only in diffractive processes, but in, for example, colour 
singlet meson exchange processes. However meson exchange gives rise to 
cross-sections which fall rapidly with increasing energy, in contrast to
diffractive processes which have constant or rising cross-sections.
Nonetheless at finite energies one may need to take account of small 
contributions from the exchange of flavour singlet meson exchange 
contributions, which can in general interfere with the dominant diffractive 
process.

Diffractive processes, so defined, are copious and varied. For example
the singly dissociative inclusive reaction
\begin{equation}
\gamma^* + p \rightarrow X + p \; \; ,
\label{DDIS}
\end{equation} 
where $X$ is an inclusive sum over hadronic states, accounts for 10-20 $\%$ 
of the $\gamma^* p$ total cross-sections at low $x$. (Here and throughout,
$\gamma^*$ indicates either a real or virtual photon, while $\gamma$ 
refers exclusively to real photons.) This reaction has stimulated an enormous 
literature already\cite{Hebecker} 
and new data will be presented here\cite{DDIS}. Of particular interest is the
behaviour for $M_X^2 \gg Q^2$,  which explores aspects of diffraction which are
not easily studied  in other processes. Exclusive processes discussed at the 
conference 
include:  elastic virtual Compton scattering
$$
\gamma^* + p \rightarrow  \gamma^* + p \; ,
$$
which is not measured directly, but is related to the $\gamma^* p$ total 
cross-sections and hence the deep inelastic structure functions via the
optical theorem; deeply virtual Compton scattering(DVCS)
\begin{equation}
\gamma^* + p \rightarrow \gamma + p \; ,
\label{DVCS}
\end{equation}
for which the first data are presented at this conference\cite{Stamen}; 
and the vector meson production processes\cite{VMP}
\begin{equation}
\gamma^* + p \rightarrow \rho + p
\label{rho}
\end{equation}
\begin{equation}
\gamma^* + p \rightarrow J/\Psi + p \; 
\label{jpsi}
\end{equation}
where measuring the vector meson decay products the enables the spin 
structure of the
interaction and the separate contributions from longtitudinal and
transverse photons to be studied. In addition, the $J/\psi$ mass introduces 
an at least
moderately  large scale into the problem even for real photons. 
Perturbative aspects of diffraction can also be enhanced by working at
high t\cite{hight} and/or by the study of diffractive jet 
production\cite{jets}. Finally some of the first  results on diffraction in 
 $\gamma^* \gamma^*$ collisions are also reported\cite{2gamma}.

\section*{Theoretical Framework}

Diffraction involves an interplay of perturbative and non-perturbative effects
which presently defies a rigorous treatment in QCD. Rather  there are 
innumerable models which throw light on different aspects of the problem
with varying degrees of success. Here we try to provide a simple framework
which can be used to classify and compare the various models and hopefully
avoid confusion. To do this we emphasize two features.

\subsection*{Vacuum exchange.} The first thing to consider is the way the 
model implements vacuum exchange. There are three main approaches, which  we 
will list for the moment and illustrate later.

\begin{itemize}

\item {\bf Regge models}, in which the vacuum exchange is usually described 
by the exchange of one or more Regge poles with vacuum quantum numbers, 
called pomerons.  

\item {\bf Gluon exchange models} in which the vacuum exchange is modelled by
the exchange of two or more gluons in a colour singlet state.

\item{\bf Quasi-optical models} in which  the projectile is regarded as a 
superposition of ``scattering eigenstates'' which are either absorbed or
scatter unchanged at fixed impact parameter on traversing the ``target.''

\end{itemize}

\subsection*{Reference frames} Different reference frames are conveniently
chosen to emphasize different aspects of the physics and caution is required in
comparing dynamical models formulated in different frames. 
Popular choices for discussing $\gamma^* p$ collisions include:  
 
\begin{itemize}

\item The {\bf infinite momentum frame} in which, for large $Q^2$ at least,
the parton distribution functions(pdfs) have a simple interpretation and
the photon is regarded as pointlike.

\item The {\bf laboratory frame} in which the incoming photon is typically
absorbed a long distance, of order $1/(M x)$ from the proton target and the
 intermediate states into which it converts are usually regarded  as 
 constituents of the photon. 

\end{itemize}

\section*{Hard and Soft Diffraction}

The study of diffraction has been transformed by the discovery of hard
diffraction in $\gamma^*p$ collisions at HERA. Here we summarize this 
discovery and some of the questions it raises.  

\subsection*{Diffraction in hadron physics}

\begin{figure}[htb]
\begin{center}
\rotatebox{270}{\psfig{file=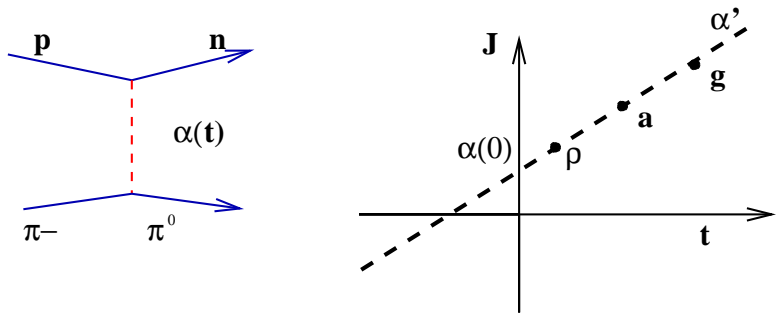,width=4cm}}
\vspace{0.2cm}
\caption{Regge pole exchange for the reaction $\pi^- p \rightarrow \pi^0 n$
and the associated meson trajectory (\ref{Mtraj})(Figure from [9]) .
\label{fig:Regge}
}
\end{center}
\end{figure}

Before discussing diffractive photoprocesses, it is useful to comment
on the ``soft diffraction'' observed in purely hadronic processes.
At high energies $s \gg t$, hadronic scattering is well-described by Regge pole
exchange, as illustrated in Figure \ref{fig:Regge} for the charge exchange reaction
$\pi^- p \rightarrow \pi^0 n$. If a single pole $i$ dominates, the 
differential cross-section for any 2 $\rightarrow$ 2 reaction satisfies
\begin{equation}
\frac{d \sigma}{d t} \propto \left(\frac{s}{s_0}\right)^{2 \alpha_i(t) - 2}  \; ,
\label{dcs}
\end{equation}
where $s_0$ is a convenient scale, usually taken to be  1 GeV$^{-2}$,
and  the {\em Regge trajectories} 
\begin{equation}
\alpha_i(t) = \alpha_i(0) +  \alpha_i^{\prime} \, t
\label{trajectory}
\end{equation}
are found to be approximately linear. They relate the observed energy 
dependence in the scattering region $t \le 0$ to the exchanged mesons
at $\alpha(t = m_j^2) = j$, where $j,m_j^2$ are the spin and mass of the 
meson respectively. The picture applies to baryon as well as meson exchange,
with an approximately universal {\em slope parameter} $\alpha^{\prime} 
\approx 1$ GeV$^{-2}$.  In contrast the {\em intercept} $\alpha_i(0)$  
depends on the flavour exchange quantum numbers $i$, with 
\begin{equation}
\alpha_M(t) \approx  0.5  +   t
\label{Mtraj}
\end{equation}
for the leading non-strange meson trajectories, leading to
cross-sections (\ref{dcs}) which fall roughly like $1/s$.

The above picture accounts remarkably well for reactions with non-zero
flavour exchange, including other features - shrinkage, factorisation,
dips - not mentioned here. It can be extended successfully to vacuum exchange 
processes by adding a single additional Regge pole to describe diffraction, 
called the pomeron. Specifically the available data on a wide range of
different reactions is consistent with the same universal trajectory\cite{DL1}
\begin{equation}
\alpha_P(t) \approx  1.08  +  0.25 t \; .
\label{SPtraj}
\end{equation}
where the  high value of the intercept $\alpha_P(0)$ reflects the fact that 
diffractive cross-sections rise slowly with energy. In addition, the pomeron
slope $\alpha_P^{\prime} \approx 0.25$ GeV$^{-2}$ differs markedly from the approximately
universal slope $\alpha_P^{\prime} \approx 1$ observed for all $q\bar{q}$
meson and $qqq$ baryon Regge poles, suggesting the pomeron is not associated
with $q\bar{q}$ meson exchange. It is rather assumed to be associated with
the exchange of gluons, so that particles lying on the pomeron trajectory
are presumably glueballs. The lightest glueball on the trajectory 
(\ref{SPtraj}) is a $2^+$ particle with a predicted mass of around
1.9 GeV. This is not unreasonable, although it must be said that little
is known from experiment about the glueball spectrum and the situation may well
be more complicated.

Finally, before leaving hadronic diffraction, we highlight two points
about Regge theory whose importance cannot be overemphasized:
 
\begin{itemize}

\item the trajectory function
(\ref{trajectory}) for any given Regge pole  depends only on $t$ and
is independent of the 
energy range and the reaction considered; and 

\item the exchange of two or more Regge poles leads to more
complicated terms - {\em Regge cuts} - which are  neglected in 
most applications, but which must be present at some level of accuracy.    

\end{itemize}

\subsection*{Diffraction in $\gamma^* p$ reactions}

The above picture of soft pomeron exchange works quite well for some real
photoprocesses like the total photoabsorption cross-section 
$\sigma_t(\gamma^*p)$ or $\rho, \omega$ or $\phi$ photoproduction.
However a  steeper rise with energy is observed if s is very large (or $x$ 
is very small) and an at least moderately hard scale enters the process.
More generally, if different data sets are parameterized by a single
Regge pole exchange formula, the intercept is found to vary roughly in the 
range
$$
 1.08 \le  \alpha_{eff}(0) \le  1.4 \; .
$$ 

\begin{figure}[htbp]
  \begin{center}
   \rotatebox{90}{\includegraphics[width=9cm,height=12cm]{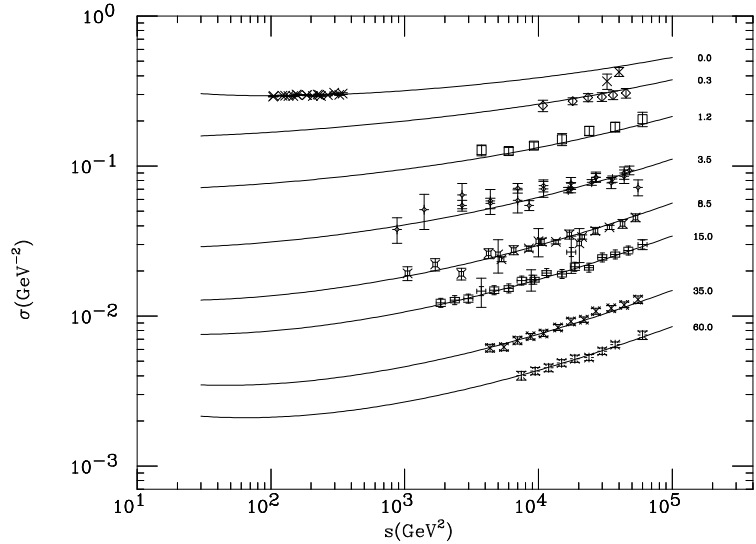}}
    \caption{ Representative sample of data points for 
the total cross-section $\sigma_{\gamma p}^{tot}$ together with curves
calculated from a colour dipole model(see below) (from [13])}
    \label{fig:sfdata}
  \end{center}
\end{figure}
\begin{figure}[hp]
\unitlength1cm
\begin{minipage}[t]{8.0cm} 
\begin{picture}(8.0,5.5) \psfig{file=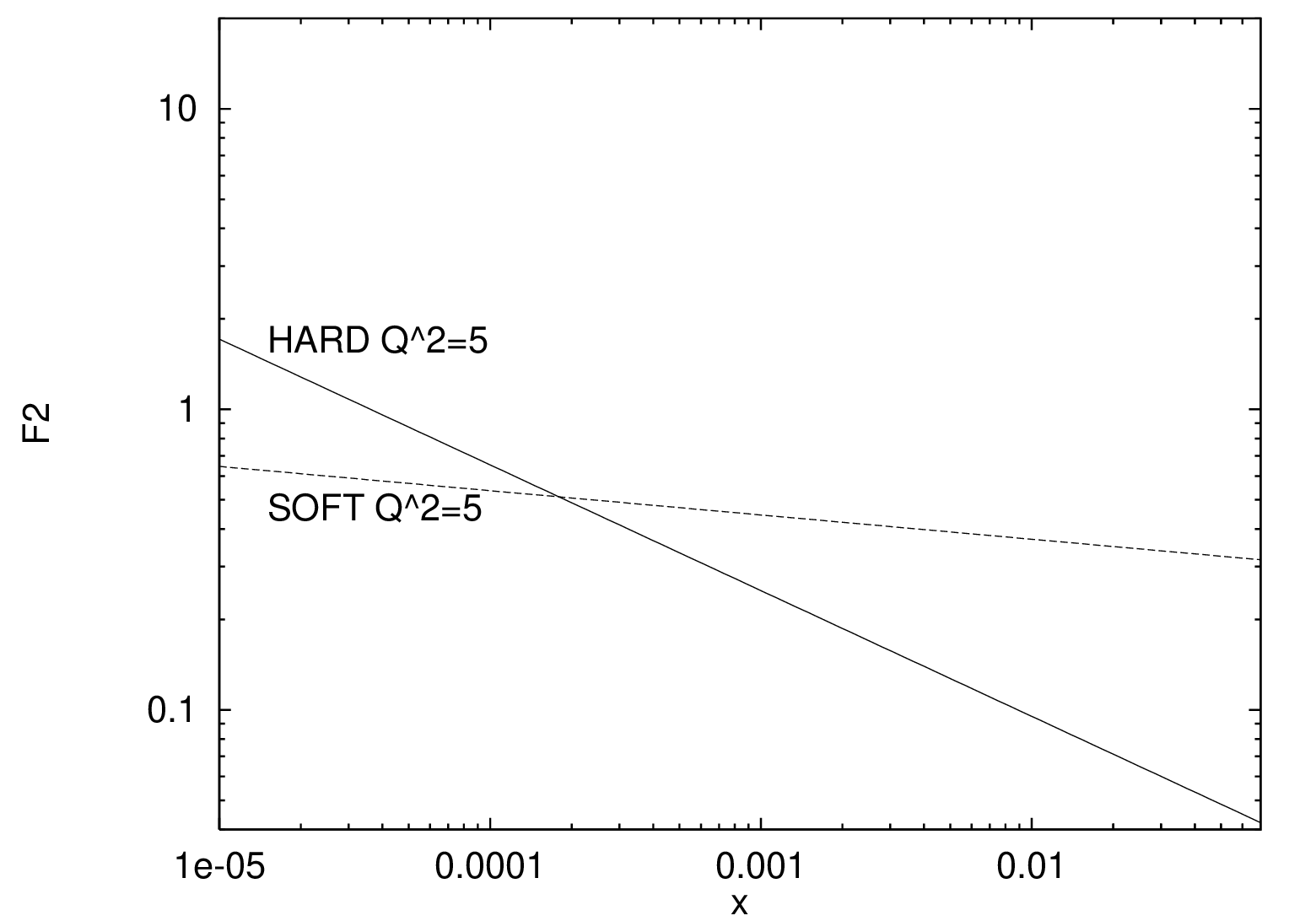,width=8cm,height=5.5cm} \end{picture} \par
\end{minipage}
\hfill
\begin{minipage}[t]{8.0cm} 
\begin{picture}(8.0,6.0) \psfig{file=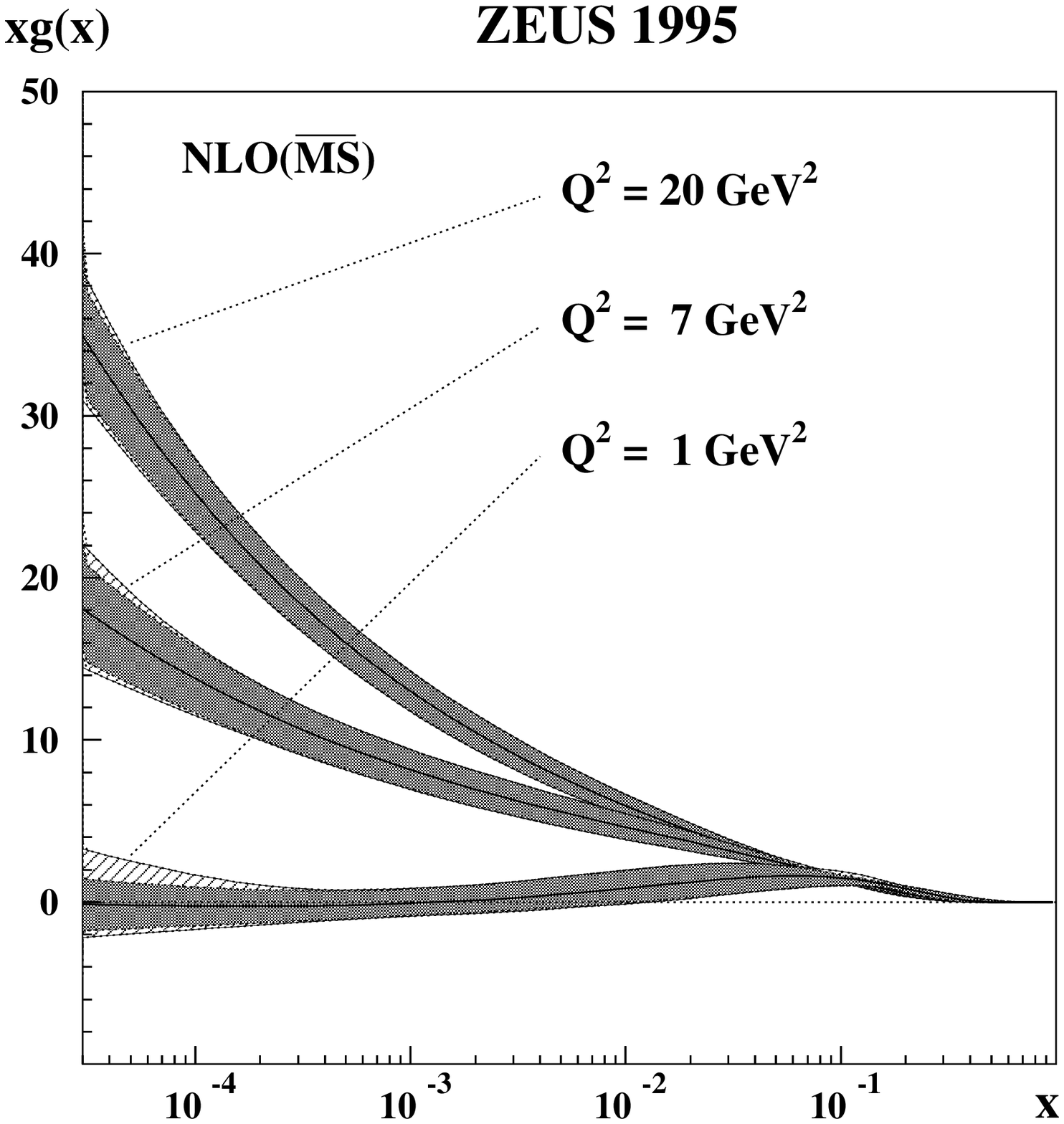,width=8cm,height=6cm}
\end{picture} 
\par
\end{minipage}
\begin{center}
\caption{Above left: relative magnitudes of hard and soft pomeron 
contributions at $Q^2=5$. As $Q^2$ increases the range of hard pomeron 
dominance extends to larger $x$. (Figure from [11])
\label{fig:f2rel}
}
\caption{Above right: the gluon density $xg(x, Q^2)$ extracted from next leading order fits
to the ZEUS $F_2(x, Q^2)$ data. (Figure from [15]) 
\label{fig:gluons}
}

\end{center}
\end{figure}

\noindent  
The particular value obtained depends on the reaction and on the ranges of 
$Q^2$ and $x$ (or equivalently $s$) considered.
This is illustrated in Figure \ref{fig:sfdata}, which shows the total 
cross-section $\sigma_{\gamma p}^{tot}$  as a function of $s$ at various 
fixed $Q^2$, where the increase in 
$ \alpha_{eff}(0)$ at high $Q^2$ and high $s$, corresponding to low $x$, is
clearly seen.

It follows from the above that  diffractive photoprocesses can not be 
described by a single Regge
pole exchange, since this requires a universal energy dependence in all cases.
The  obvious interpretation is that there is a new phenomenon
- ``hard diffraction'' - which becomes dominant for hard enough scales
and large enough energies. If one assumes that this can also be approximated
by a Regge pole one is led to the
hypothesis of two pomerons: the soft pomeron (\ref{SPtraj}) which dominates
in hadronic diffraction and some ``soft'' photoprocesses; and a second
``hard pomeron'' which dominates for hard enough scales and large enough 
energies. This hypothesis has been explored by Donnachie and
Landshoff\cite{DL2} who obtain an excellent fit to data on the proton
structure function, the charmed structure function and on $ J/\psi$
production (\ref{jpsi}) for a hard pomeron trajectory
$$
\alpha_P(t) \approx  1.42  +  0.10 t \; .
$$
The varying energy dependence arises from the varying relative importance of 
the two contributions, which is illustrated in Figure \ref{fig:f2rel} for the 
proton structure function. Alternatively, Gotsman will discuss\cite{gotsman} 
a two component model in which the ``hard component'' is described by a model 
based on perturbative QCD. In both cases, between the regions of soft and hard 
diffraction at $Q^2 = 0$ and  high $Q^2$ respectively, there is an
extensive transition region in which both can be important.

\subsection*{Hard diffraction and QCD}

The discovery of hard diffraction at HERA opens up the subject 
to perturbative methods in QCD. These can be  based on resummations
of the perturbative expansion retaining leading terms 
in  $\ln( Q^2/\Lambda^2)$ (DGLAP) or leading terms in 
$\ln(1/x)$ (BFKL), since both these quantities become large in the 
relevant kinematic region. The former is the most familiar and has it's
best known application in the DGLAP evolution of the structure functions.
In particular, it has long been known\cite{derujula} that this can generate 
an increasingly steep low-$x$ behaviour as $Q^2$ increases, and a good fit 
to the data can be obtained for $Q^2 \ge 1$ GeV$^2$ with the gluon 
distributions $xg(x,Q^2)$ shown in Figure \ref{fig:gluons} \cite{ZEUSfit}.
The gluon densities at $Q^2 \approx 1$ GeV$^2$ are unstable, implying that the
DGLAP picture can not be trusted at such low $Q^2$ values, but the success at
higher $Q^2$ values is impressive. This ``DGLAP picture,'' based on the 
dominance of gluon ladder exchanges(see e.g. the talk of 
Gotsman\cite{gotsman}),
can be extended to other diffractive processes at high $Q^2$, especially
for those processes where, in lowest order at least, one can prove 
factorisation  into terms describing the fluctuation of the initial photon
into $q \bar{q}$ pairs; the formation of the final particle from the said
pairs; and the interaction of the $q \bar{q}$ with the 
proton \cite{bib:fact}. However the gluon distributions required 
are ``skewed''
parton distributions\cite{bib:skew}, which take into account the fact 
that the incoming
and outgoing protons in inelastic processes like (\ref{DDIS}, \ref{DVCS},
 \ref{rho}, \ref{jpsi})
have different momenta, even in the forward direction. The empirical study of 
skewed parton corrections has only just begun\cite{Stamen} \cite{bib:ffs}.


A potential problem with  leading, next leading .. $\ln Q^2$ approximations
is that as $x \rightarrow 0$, neglected
terms might  become important because although they are lower order 
in $\ln Q^2$,
they are leading order in  $\ln (1/x)$. Thus
one might expect to see a breakdown of DGLAP at very small-$x$ - but how 
small? This gives rise to the alternative  BFKL approach of 
leading $\ln (1/x)$ resummation. In leading order this approach  
predicted the hard pomeron intercept with apparent success, but it runs into 
serious difficulties beyond leading
order. This topic will be discussed by Ross\cite{bib:ross} while a succinct
comparison of the Regge, BFKL and DGLAP approaches and their relation to
each other may be found in the recent review of Ball and 
Landshoff\cite{bib:bl}.   

In the rest of this talk we will concentrate on two more phenomenological
 questions:

\begin{itemize}

\item Can we find a unified description of both hard and soft diffraction and
of the wide variety of diffractive processes? 

\item When can we expect to see  so-called ``gluon saturation'' effects 
at small $x$?

\end{itemize}

These are conveniently addressed in the context of the
 {\em colour dipole model}, to which we
immediately turn.

\section*{The colour dipole model}

\begin{figure}[htb]
\begin{center}
\psfig{file=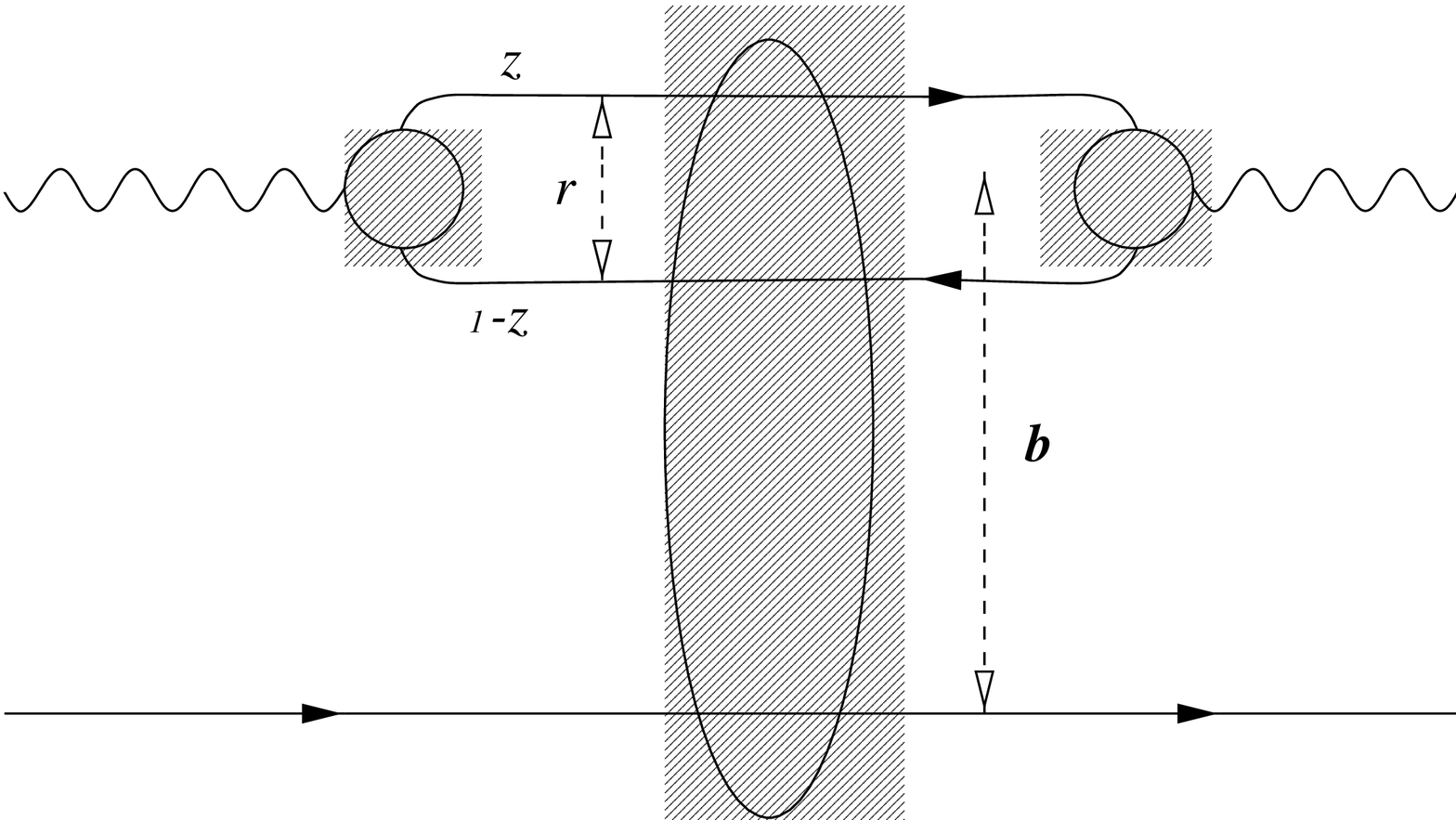,width=10cm,height=3cm}
\caption{The colour dipole model for $\gamma^* p \to \gamma^* p$.
\label{fig:fluct}
}
\end{center}
\end{figure}

Singly dissociative diffractive $\gamma$ p processes are conveniently 
described in the rest frame of the hadron using a picture in which the 
incoming photon intially dissociates  into a $q \bar{q}$ pair
a long distance - typically of order of the ``coherence length'' $1/Mx$ 
- from the target proton.  Assuming that the resulting partonic/hadronic 
state evolves slowly compared to the size of the proton or nuclear target,
it can be regarded as frozen during the interaction. In this approximation,
the process will factorize into a probability for the photon to have evolved 
into a given state $ | \alpha \! >$, times the amplitude for that state to 
interact with the target.
In the colour dipole model, ~\cite{bib:dcs_nik,bib:dip_muel_1}
the dominant states $ | \alpha \! >$  are assumed to 
be $q \bar{q}$ states of given transverse size. Specifically 
\begin{equation}
  \label{eq:photon_wf}
  |\gamma\rangle = \int \mbox{d}z \mbox{d}^{2}r \ \psi (z,r) |z,r\rangle + 
\ldots \; ,
\end{equation}
where $r$ is the transverse size of the pair, $z$ is the fraction of light 
cone energy carried by the quark and $\psi (z,r)$ is the \emph{light cone 
wave function} of the photon. Assuming that these states are  scattering
eigenstates (i.e. that $z, r$ remain unchanged in diffractive scattering) 
the elastic scattering amplitude for $\gamma^* p \to \gamma^* p$ is 
specified by Figure \ref{fig:fluct}. This leads via the optical theorem to  
\begin{equation}
  \label{eq:sigma_tot}
   \sigma^{\gamma^{*}p}_{T,L} = \int \mbox{d}z \mbox{d}^{2}r \ |\psi_{\gamma}^
{T,L}(z,r)|^{2} \sigma(s,r,z) \; , 
\end{equation}
for the $\gamma^* p$ total cross-section in deep inelastic scattering,
where  $\sigma(s,r,z)$ is the total cross-section
for scattering dipoles of specified  $(z,r)$  from a proton at 
fixed $ s = W^2$. This ``dipole cross-section'' is a universal quantity for 
singly-dissociative 
diffractive  processes on a proton target, playing a similarly 
fundamental role in, for example, open diffraction (\ref{DDIS}), exclusive
 vector meson production (\ref{rho}) and (\ref{jpsi}) and deeply virtual 
Compton scattering (\ref{DVCS}).

The dipole cross-section has been evaluated by several 
groups\cite{amirim}. Although the assumptions made to do this vary, there are
some features in common. The dipole cross-section at a given energy is assumed
to be approximately  ``geometrical'', i.e. to depend on the transverse size 
 $r$ of the dipole, but not to depend on $z$. In addition, approximate QCD 
behaviour(colour transparency) for small dipoles $r \rightarrow 0$ and 
``hadronic behaviour''

\begin{figure}[htb]
\begin{center}
\psfig{file=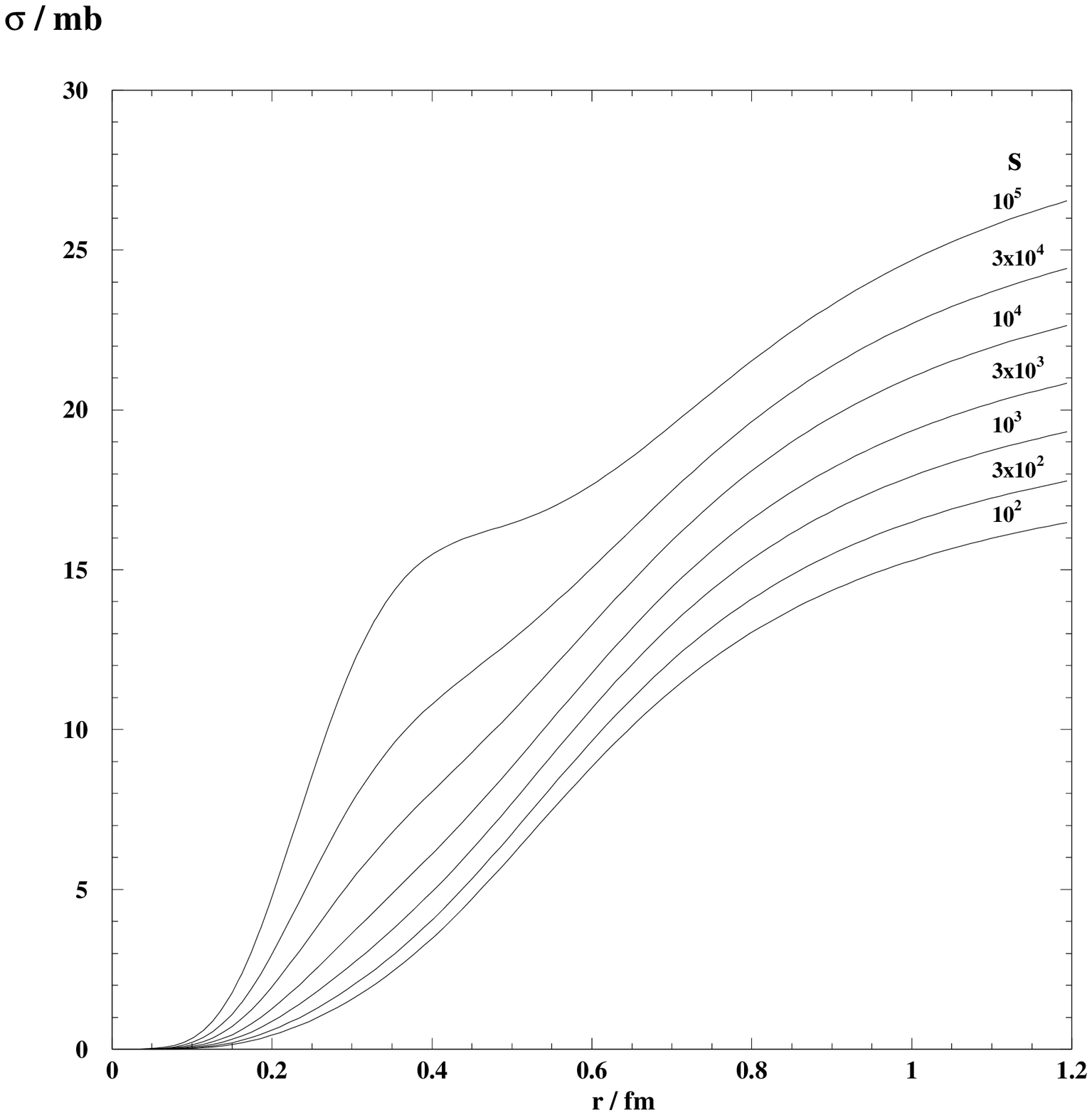, width=12cm,height=7cm}
\caption{The dipole cross-section as a function of s in the HERA range.
\label{fig:dcs}
}
\end{center}
\end{figure}

\noindent
for  large dipoles $r \approx 1$fm are incorporated in varying degrees of 
detail\footnote{Very large dipoles $r \gg 1$fm make a negligible contribution,
since the wavefunction factor in (\ref{eq:sigma_tot}) decreases exponentially 
at large $r$.}.
A useful summary and comparison of the various approaches may be found in the
 recent review of McDermott\cite{amirim}.
From now  on I shall present results from Forshaw, Kerley and 
Shaw\cite{bib:our1}, \cite{bib:our2} - \cite{bib:our4} who  
have extracted the dipole cross-section
from DIS and real photoabsorption data assuming a form with two 
terms with a  Regge type $s$ dependence:
\begin{equation}
\label{gammatot}
  \sigma(s,r)  =   a_{soft}(r) s^{\lambda_{S}} + a_{hard}(r) s^{\lambda_{H}} 
\; \end{equation}
where the values $\lambda_S \approx 0.08$, $\lambda_H \approx 0.42$
resulting from the fit are characteristic of the soft and hard pomeron
respectively. The functions $a_{soft}(r)$, $a_{hard}(r)$ are chosen so
that for  small dipoles the hard term dominates yielding a 
behaviour $\sigma \rightarrow r^2 (r^2 s)^{\lambda_H}$ as $r \rightarrow 0$
in accordance with colour transparency ideas; while for large dipoles
 $r \approx 1$ fm the soft 
term dominates with a hadronlike behaviour $\sigma \approx \sigma_0 
(r^2 s)^{\lambda_S}$. Correspondingly the photon wavefunction is assumed to be
perturbative for small dipoles, with a simple ansatz for confinement
effects at large $r$. The resulting 
dipole cross-section, determined from DIS and real photoabsorption data, 
is shown in Figure \ref{fig:dcs} for various energies in the HERA region.

\begin{figure}[htb]
\begin{center}
\psfig{file=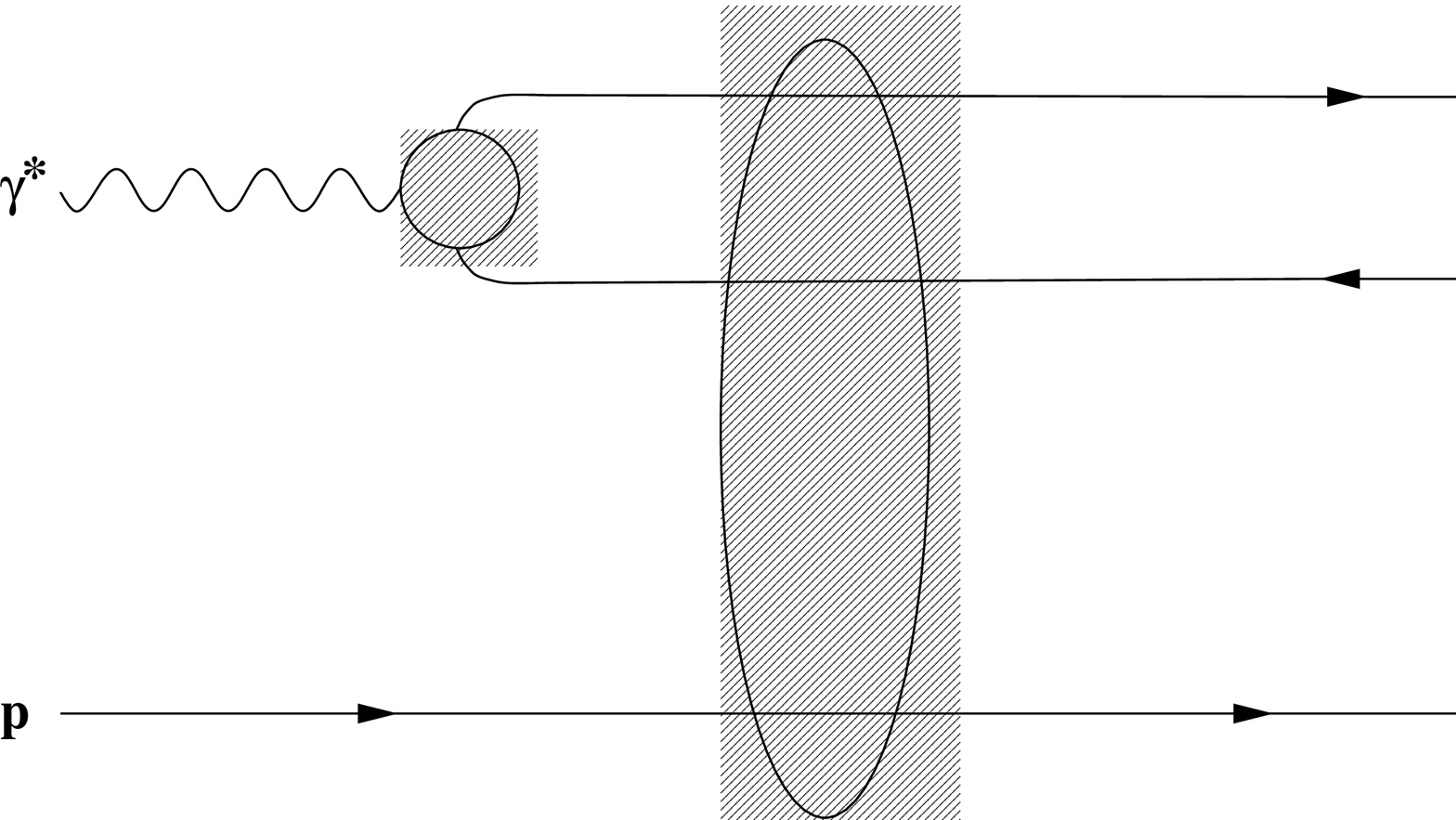, width=8cm,height=3cm}
\vspace{0.2cm}
\caption{The dipole contribution to open diffraction (\ref{DDIS}).Figure from
[25].
\label{fig:DDIS}
}
\end{center}
\end{figure}

The above dipole cross-section, determined from DIS and real 
photoabsorption data, can be used to predict results for other diffractive 
processes. Successful predictions have been obtained for:

\begin{itemize}

\item  the charmed
structure function\cite{bib:our1}  by retaining only the charmed quark
loop in Figure \ref{fig:fluct}; 

\item open diffraction (\ref{DDIS}) from
Figure \ref{fig:DDIS}, together with an additional contribution 
from intermediate  $ q \bar{q} g$ states  which is important for 
large diffractive masses $m_X^2 \gg Q^2$, but  small elsewhere  
\cite{bib:our2}.

\item  virtual Compton scattering (\ref{DVCS}), by replacing
the final state photon in Figure \ref{fig:fluct} by a real 
photon \cite{bib:our4}.

\end{itemize}

The same dipole cross-section can also be used to predict vector meson
production reactions like (\ref{rho}, \ref{jpsi}), but in this case
the vector meson wavefunctions are also required.

\subsection*{Saturation}

The dipole model is particularly useful for discussing saturation effects, 
since it incorporates both soft and hard diffraction, associated with small
and large dipoles respectively. There are actually two  types of 
saturation effect, which are quite distinct and should not be confused.

{\bf Low $Q^2$ saturation.}
As can be seen in Figure \ref{fig:dcs}, the dipole cross-section increases
rapidly as a function of the dipole size $r$ at small $r$, but then 
``saturates'' to a slowly
varying cross-section of hadronic size at larger $r$ values. This change
- and the fact that it shifts to smaller $r$ as $s$  increases -is crucial to 
describe the form of the change from approximate scaling  to 
the observed $Q^2 \rightarrow 0$  (and hence $x \rightarrow 0$) behaviour
at fixed $s$. To see this we note that the  $Q^2$ dependence in 
(\ref{eq:sigma_tot}) arises entirely from the wavefunction. As $Q^2$ decreases,
larger $r$ values are explored and the slowly varying dipole cross-section
results in a weakening $Q^2$ dependence for  $\sigma_{\gamma* p}$.  
When  $Q^2 \ll 4 m_q^2$, where $m_q$ is the 
constituent quark mass, the wavefunction and $\sigma_{\gamma* p}$ become 
independent of $Q^2$  so 
that $F_2 \propto Q^2$ as $Q^2 \rightarrow 0$ as required.

\begin{figure}
\begin{center}
\psfig{file=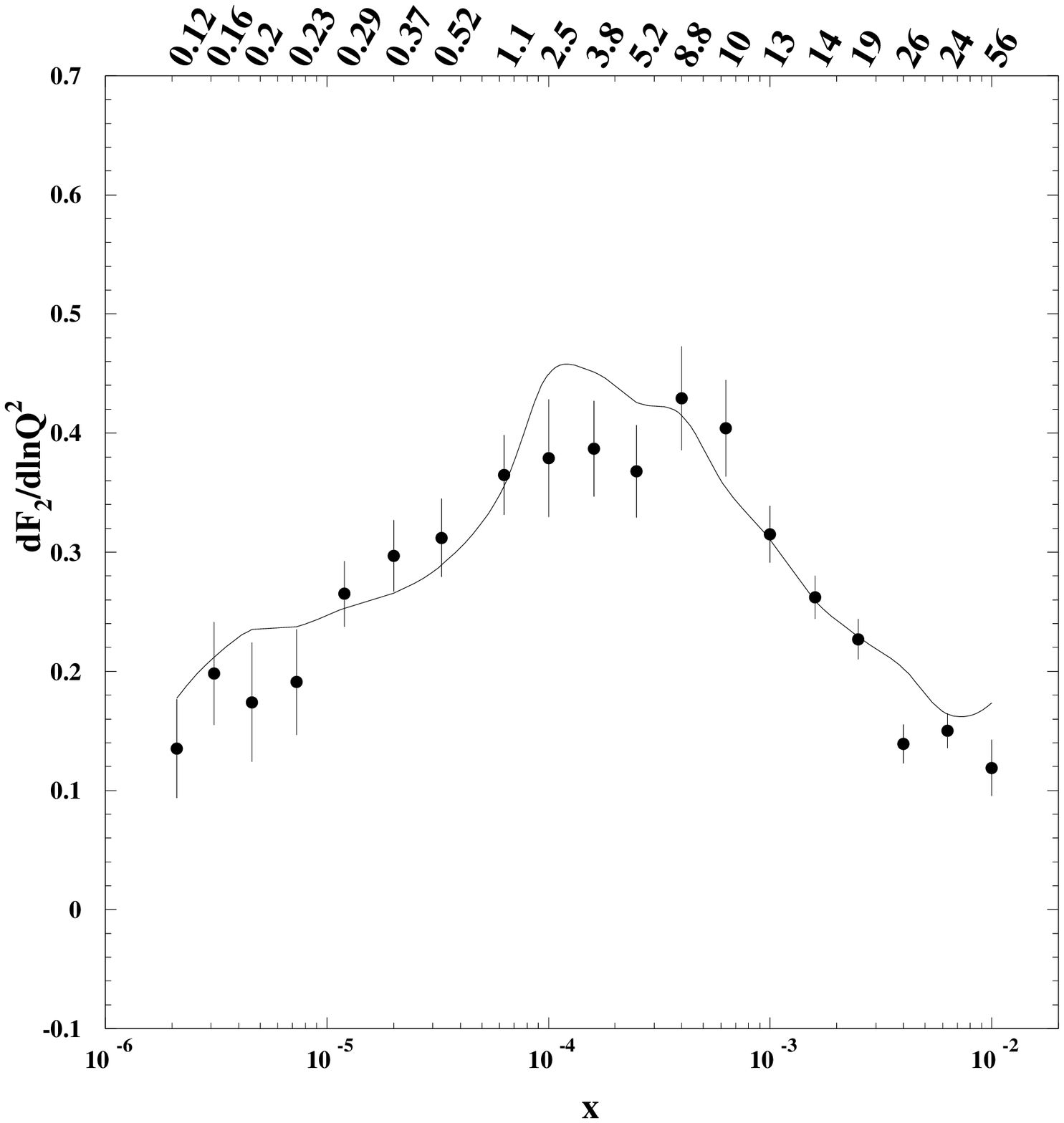, width=11cm,height=6.8cm}
\caption{The Caldwell plot. Predictions are made at the $Q^2$ values of
the data points and roughly interpolated. (Figure from [25])
\label{fig:caldwell}
}
\end{center}
\end{figure}

{\bf Gluon saturation.}
For high enough energies, the assumed $s^\lambda \;( \lambda > 0)$ behaviours 
assumed above  must be tamed by unitarity effects, especially for the hard
term with $ \lambda_H \approx 0.4$. At fixed $Q^2$, $x \rightarrow 0$ as
$s \rightarrow \infty$ and the resulting  softening of the corresponding 
$x^{- \lambda_H}$ behaviour is associated with gluon saturation in the
quark-parton language. Gluon saturation can be incorporated into dipole and 
other closely related models
by hand \cite{bib:gbw,bib:mcd} or using the eikonal approximation
\cite{bib:levin,bib:capella}  but are not included
in (\ref{gammatot}). Hence the fact that  an excellent fit is obtained
to the DIS data using (\ref{gammatot})  means that the current HERA data are
not at sufficiently high $s$ to {\em require} the saturation effects that are
built into some other  dipole models ~\cite{bib:gbw,bib:mcd}.
We note that our model agrees
with the standard Caldwell plot Figure \ref{fig:caldwell}, where the turn 
over as $x$
decreases occurs because $Q^2$ is also decreasing and is understood as 
a low $Q^2$ saturation effect. No such effect is predicted
in our model if $x$ is decreased at fixed $Q^2$, as confirmed by the
preliminary  ZEUS97 data\CITE{ZEUS97}.

A strong indication of when saturation effects will be needed is given
in Figure \ref{fig:dcs}. As can be seen, the cross-section for small dipoles
is initially small but increases rapidly and  at the top of 
the accessable HERA range is becoming commensurate with the
slowly increasing ``hadronic'' behaviour af the large dipoles. It is at this
point that saturation effects are expected to become important;  if they 
don't, the cross-section for small dipoles will exceed that for large dipoles 
at higher energies and the dipole cross-section will paradoxically decrease 
with increasing size $r$. Saturation effects are therefore expected to play
an important role just beyond beyond the HERA range, in the planned THERA 
region with $s_{max} \approx 10^6$ GeV$^2$.

\end{document}